\documentclass[superscriptaddress,secnumarabic,nobibnotes,aps,prd,showpacs,nofootinbib]{revtex4}
\usepackage{graphicx}
\usepackage{epsf}
\usepackage{bm}
\usepackage{amsmath}
\usepackage{amsfonts}
\usepackage{amssymb}
\usepackage{epstopdf}
\usepackage{natbib}
\usepackage{color}

\providecommand{\U}[1]{\protect\rule{.1in}{.1in}}
\newcommand{\be}{\begin{equation}}
\newcommand{\ee}{\end{equation}}

\newcommand{\mincir}{\raise
-3.truept\hbox{\rlap{\hbox{$\sim$}}\raise4.truept\hbox{$<$}\ }}
\newcommand{\magcir}{\raise
-3.truept\hbox{\rlap{\hbox{$\sim$}}\raise4.truept\hbox{$>$}\ }}

\input{tcilatex}

\begin{document}

\title{Two-fluid solutions of particle-creation cosmologies}
\author{Supriya Pan}
\email{supriya.maths@presiuniv.ac.in}
\affiliation{Department of Mathematics, Presidency University, 86/1 College Street,
Kolkata 700073, India}
\author{John D. Barrow}
\email{J.D.Barrow@damtp.cam.ac.uk}
\affiliation{DAMTP, Centre for Mathematical Sciences, University of Cambridge,
Wilberforce Rd., Cambridge CB3 0WA, UK}
\author{Andronikos Paliathanasis}
\email{anpaliat@phys.uoa.gr}
\affiliation{Instituto de Ciencias F\'{\i}sicas y Matem\'{a}ticas, Universidad Austral de
Chile, Valdivia, Chile}
\affiliation{Institute of Systems Science, Durban University of Technology, PO Box 1334,
Durban 4000, Republic of South Africa}

\begin{abstract}
Cosmological evolution driven incorporating continuous particle creation by
the time-varying gravitational field is investigated. We consider a
spatially flat, homogeneous and isotropic universe with two matter fluids in
the context of general relativity. One fluid is endowed with gravitationally
induced \textquotedblleft adiabatic\textquotedblright\ particle creation,
while the second fluid simply satisfies the conservation of energy. We show
that the dynamics of the two fluids is entirely controlled by a single
nonlinear differential equation involving the particle creation rate, $%
\Gamma (t)$. We consider a very general particle creation rate, $\Gamma (t)$%
, that reduces to several special cases of cosmological interest, including $%
\Gamma =$ constant, $\Gamma \propto 1/H^{n}$ ($n\in \mathbb{N}$), $\Gamma
\propto \exp (1/H)$. Finally, we present singular algebraic solutions of the
gravitational field equations for the two-fluid particle creation models and
discuss their stability.
\end{abstract}

\maketitle

\section{Introduction}

Current astronomical observations show that the recent history of the
universe is consistent with accelerated expansion that might be described
either by some exotic dark energy fluid in the framework of Einstein's
gravitational theory or by a modification of the gravitational theory
itself. This state of affairs has motivated the scientific community to look
for alternative theories which can reproduce the effects of the dark energy
or modified gravity models in a natural way. The theory of gravitational
particle production has a long history. The production of quantum particles
by the gravitational field was first investigated in the context of the
early universe \cite{aniso} as a device for damping initial anisotropies,
following Schr\"{o}dinger's first look at quantum fields in an expanding
isotropic universe \cite{Sch1939}, although such damping was likely to
produce far more radiation entropy than is observed in the microwave
background \cite{BMatz}. Later, a classical counterpart to the quantum
particle production was described by Grishchuk \cite{grish}, and
subsequently, cosmological particle production was examined extensively in
the literature and used as a means of describing the generation of
inhomogeneities during inflation. Many investigators \cite{Prigogine-inf,
Abramo:1996ip, Gunzig:1997tk,Zimdahl:1999tn, SSL09, LJO10, LB10, JOBL11,
LBC12, LGPB14, CPS14, FPP14, NP15, lima16, NP16, HP16, RCN16, PHPS16} (and
references therein), have argued that the particle production process might
be considered as another approach to explain different phases of the
universe's evolution. In particular, it has been found that such particle
creation phenomena can describe early acceleration \cite{Prigogine-inf,
Abramo:1996ip, Gunzig:1997tk, Zimdahl:1999tn, RCN16} and late acceleration
in the expansion of the universe \cite{SSL09, LJO10, LB10, JOBL11, LBC12,
LGPB14, CPS14, FPP14, NP15, lima16, NP16, HP16, PHPS16}\emph{.} Particle
creation might provide a way to unify the early- and late- accelerated
expansions with intermediate radiation and matter dominated phases of the
universe \cite{LBC12, PHPS16}. The development of a theory of particle
production rests upon the choice of the creation rate $\Gamma (t)$, which is
an unknown function of time and it can only be determined from the quantum
field theory (QFT) in curved spacetime; however, QFT is not yet able to
provide the exact functional form for $\Gamma (t)$. Therefore, it is
convenient to study some different particle creation rates and build up a
picture of the classes of cosmological evolution that arise from different
choices for $\Gamma (t)$. This approach is phenomenological but it allows us
to constrain the choice of the particle creation rates from their effects.
It is possible to identify the particle creation rates with the dark energy
equation of state, or the modified gravity effects, because the introduction
of a particle creation rate is equivalent to a time-varying (effective)
equation of state \cite{bar01}. Following this, several phenomenological
choices for the form of $\Gamma (t)$ have already made by different authors 
\cite{Prigogine-inf, Abramo:1996ip, Gunzig:1997tk,Zimdahl:1999tn, RCN16,
SSL09, LJO10, LB10, JOBL11, LBC12, LGPB14, CPS14, FPP14, NP15, lima16, NP16,
HP16, PHPS16}. However, a theory for a general form of $\Gamma (t),$
recovering some well known choices as special cases, is an appealing goal.

It is useful to elaborate on the connection between the quantum particle
production studied in references like \cite{aniso} and the classic picture
of particle production that we use here by introducing a non-adiabatic term
in the conservation equation for one of our fluids. In effect, this is
equivalent to endowing that fluid with a bulk viscosity \cite{bulk}, which
is the only form of dissipative stress allowed in the isotropic models.
There are bulk viscous cosmological solutions that display an increasing
density of particles for a period, starting from zero at an initial
curvature singularity, before their density begins to fall adiabatically
because of the domination of the expansion effects. These effects lead to
entropy increase so long as the bulk viscous coefficient is positive \cite%
{bulk}. Our model described in eq. (\ref{cons1}) below is a cosmology with a
classical bulk viscous stress that can induce particle creation in the form
of particle density increases.

We will consider a spatially flat homogeneous and isotropic universe where
the gravitational sector is described by general relativity and the total
matter sector is divided into two fluids where one fluid (named as fluid I)
is endowed with \textquotedblleft adiabatic\textquotedblright\ particle
production process and the second fluid (fluid II) is independently
conserved. With this set up, we find that the dynamics of the universe is
governed by a nonlinear differential equation which is dependent on the
choice of the particle creation rate. We explore the dynamical evolution of
this cosmological model using exact solutions of Einstein's equations for a
generalized choice of the particle creation rate, $\Gamma (t)$. By applying
the method of singularity analysis to the nonlinear differential equations 
\cite{cotsakis1,cotsakis2,cotsakis3,paperleach, anleach,
Paliathanasis:2016vsw}, we find that their solution can be written in terms
of the Laurent series around their movable singularity.

The work has been organized in the following way. In section \ref{sec:ede},
we describe the field equations for a two-fluid system endowed with a
particle creation mechanism in the framework of Einstein gravity. In section %
\ref{sec-analytic} we describe the analytic solutions for the cosmological
model with a general series form for the particle creation rate. Finally, in
section \ref{discu} we summarise our main findings.

\section{Field equations for a two-fluid model with particle creation}

\label{sec:ede}

We consider the spatially flat Friedmann-Lema\^{\i}tre-Robertson-Walker
(FLRW) metric,

\begin{equation*}
ds^{2}=-dt^{2}+a^{2}(t)\left[ dr^{2}+r^{2}\left( d\theta ^{2}+\sin
^{2}\theta d\phi ^{2}\right) \right] ,
\end{equation*}%
where $a(t)$ is the expansion scale factor and $t$ is comoving proper time.

We assume that the gravitational sector is described by the Einstein gravity
and the matter sector is described by the two-fluid system, with energy
densities and pressures: $(\rho ,p)$ (fluid I) and $(\rho _{1},p_{1})$
(fluid II) where one fluid (fluid I) is endowed with \textquotedblleft
adiabatic\textquotedblright\ particle creation mechanism. With the term
\textquotedblleft adiabatic\textquotedblright\ we mean that the entropy per
particle remains constant.

The gravitational equations for such two fluid system can be written as (in
the units $8\pi G=1$)

\begin{eqnarray}
\frac{\dot{a}^{2}}{a^{2}} &=&\frac{\rho +\rho _{1}}{3},  \label{efe1} \\
2\frac{\ddot{a}}{a}+\frac{\dot{a}^{2}}{a^{2}} &=&-(p+p_{c}+p_{1}),
\label{efe2}
\end{eqnarray}%
where an overhead dot represents the cosmic time differentiation; $p$, $\rho 
$ are respectively the thermodynamic pressure and the energy density for
fluid I with $p=(\gamma -1)\rho $, ($0<\gamma \leq 2$ is the barotropic
state parameter), while $p_{1}$, $\rho _{1}$ represent the same quantities
for fluid II with $p_{1}=(\gamma _{1}-1)\rho _{1}$, ($0<\gamma _{1}\leq 2$
is the barotropic index of fluid II). The quantity $p_{c}$ is the creation
pressure due to the production of particles and this is related to fluid I by

\begin{equation}
p_{c}=-\frac{\Gamma (t)}{3H}(p+\rho ),  \label{efe3}
\end{equation}%
where $H=\dot{a}/a$ is the Hubble rate of the FLRW universe and $\Gamma \geq
0$ is the rate of particle creation, which is an unknown quantity. The exact
functional form for $\Gamma (t)$ can only be determined from the quantum
field theory in curved spacetimes, otherwise it must be modelled by a
general functional dependence on physical quantities. Since that subject is
not fully developed yet, we assume specific functional forms for $\Gamma $
and try to explain the ensuing cosmological evolution. An important
observation about the pressure term (\ref{efe3}) is that if fluid I
describes the cosmological constant, with $p=-\rho $, then $p_{c}$\
vanishes. 

We note that the expression (\ref{efe3}) follows from the Gibbs equation
together with the property that the entropy per particle is constant. In
particular, if $n$\ stands for the particle number density, then the
nonconservation of fluid I particles follows%
\begin{equation}
\dot{n}+3Hn=n\Gamma \left( t\right) ,
\end{equation}%
while the Gibbs equation,  $Tds=d\left( \frac{\rho }{n}\right) +pd\left( 
\frac{1}{n}\right) ,$\ gives%
\begin{equation}
nT\dot{s}=\dot{\rho}+3H\left( \left( \rho +p\right) +p_{c}\right) ,
\end{equation}%
in which $p_{c}$\ is defined \ through equation (\ref{efe3}) and  `$s$'\
denotes the entropy per particle which has been considered to be constant.
Hence, for the fluid I one finally has%
\begin{equation}
\dot{\rho}+3H\left( 1-\frac{\Gamma (t)}{3H}\right) (p+\rho
)=0\Leftrightarrow \dot{\rho}+3\gamma H\rho =\Gamma (t)\gamma \rho \;,
\label{cons1}
\end{equation}%
from which one can solve the evolution for $\rho $ \cite{bar01}: 
\begin{equation}
\rho =\frac{\rho _{A}}{a^{3\gamma }}\exp \left( \gamma \int \Gamma
(t)dt\right)   \label{evol-rho}
\end{equation}%
where $\rho _{A}>0$ is an integration constant. As already noted in \cite%
{bar01}, the evolution equation for $\rho $ in (\ref{evol-rho}) actually
provides an equivalent description for a fluid with a dynamical equation of
state. We can interpret the first of (\ref{cons1}) as the conservation for a
fluid with a bulk viscous coefficient equal to $\Gamma (t)\gamma \rho /9H^{2}
$, \cite{bulk}. For different creation rates, different fluids can be
realized for a fixed $\gamma $. The second fluid, (fluid II), is assumed not
to have any interaction with fluid I, and so obeys the usual conservation
equation: 
\begin{equation}
\dot{\rho}_{1}+3H(p_{1}+\rho _{1})=0\ \Leftrightarrow \rho _{1}=\rho
_{1,0}\;a^{-3\gamma _{1}},  \label{cons2}
\end{equation}%
where the integration constant $\rho _{1,0}$ denotes the present value of $%
\rho _{1}(t)$.

Using the field equations (\ref{efe1}), (\ref{efe2}), together with the
evolution of the second fluid from (\ref{cons2}), we can derive the master
differential equation for this two-fluid system, as {%
\begin{equation}
2\dot{a}\ddot{a}+\left( 3\gamma -2\right) \frac{\dot{a}^{3}}{a}-\rho
_{1,0}(\gamma -\gamma _{1})a^{1-3\gamma _{1}}\,\dot{a}-\gamma \Gamma \left( 
\dot{a}^{2}-\frac{\rho _{1,0}}{3}a^{2-3\gamma _{1}}\right) =0,
\label{mas.01}
\end{equation}%
where, }as already mentioned, $\gamma ,\gamma _{1}>0$. We remark that in the
limit $\gamma _{1}=\frac{2}{3}$, the dynamics reduces to the previous work
with curvature of the universe added, see \cite{bar01}.

We continue with the definitions of the total equation of state $w_{\mathrm{%
tot}}$ and the deceleration parameter $q(a)$ providing a clear picture of
the different phases of the universe. The total equation of state $w_{%
\mathrm{tot}}$ of the two-fluid system is defined by $w_{\mathrm{tot}}=P_{%
\mathrm{tot}}/\rho _{\mathrm{tot}}$, where $P_{\mathrm{tot}}=p+p_{c}+p_{1}$,
and $\rho _{\mathrm{tot}}=\rho +\rho _{1}$. Using the field equations (\ref%
{efe1}) and (\ref{efe2}), this can be recast into the following simplified
expression

\begin{equation}
w_{\mathrm{tot}}=-1+\frac{1}{3H^{2}}\Bigg[\gamma _{1}\rho _{1}+\gamma \left(
1-\frac{\Gamma }{3H}\right) (3H^{2}-\rho _{1})\Bigg]=-1+\frac{1}{\rho +\rho
_{1}}\Bigg[\gamma _{1}\rho _{1}+\gamma \rho \left( 1-\frac{\Gamma }{\sqrt{%
3(\rho +\rho _{1})}}\right) \Bigg],  \label{tot-eos}
\end{equation}%
where $\rho _{1}$ can be found from (\ref{cons2}).

We note from the total equation of state in (\ref{tot-eos}), that we can
derive different bounds on it. In other words, the total fluid could behave
like a radiation fluid ($w_{\mathrm{tot}}=1/3$), or a dust fluid ($w_{%
\mathrm{tot}}=0$), or a pure cosmological constant ($w_{\mathrm{tot}}=-1$),
depending on the particular different particle creation rate $\Gamma ,$
independent of $\gamma $ and $\gamma _{1}$. In Table \ref%
{tab:characterization} we have explicitly listed different particle creation
rates corresponding to distinct phases of the universe's evolution.

Following the above equations, we can see that the rate of particle
creation, $\Gamma $, plays a key role in determining different stages of the
universe's evolution. In addition, it is interesting to calculate the
particle creation rate at the transition scale factor $a=a_{t}$ by solving
the equation $q(a=a_{t})=0$, which gives,

\begin{equation*}
\Gamma (a=a_{t})=\frac{3H}{\gamma (3H^{2}-\rho _{1})}\Bigl[(3\gamma
-2)H^{2}+(\gamma _{1}-\gamma )\rho _{1}\Bigr].
\end{equation*}

In the next section, we introduce a generalized model for $\Gamma $ and
apply the singularity test \cite{paperleach, anleach, Paliathanasis:2016vsw}
to find the exact analytical solutions for the master equation (\ref{mas.01}%
).

\begingroup

\begin{center}
\begin{table*}[tbp]
\begin{tabular}{ccccc}
\hline\hline
Mimicking Fluid & ~~~ Total equation of state ($w_{\mathrm{tot}}$) & ~~~
Particle creation rate ($\Gamma$) &  &  \\ \hline
Radiation & $w_{\mathrm{tot}} = 1/3$ & $\Gamma = 3 H \left[1 - \frac{4H^2 -
\gamma_1 \rho_1}{\gamma (3H^2 - \rho_1)}\right]$ &  &  \\ 
Dust & $w_{\mathrm{tot}} = 0$ & $\Gamma = 3 H \left[1 - \frac{3H^2 -
\gamma_1 \rho_1}{\gamma (3H^2 - \rho_1)}\right]$ &  &  \\ 
Cosmological constant & $w_{\mathrm{tot}} = -1$ & $\Gamma = 3 H \left[ 1 + 
\frac{\gamma_1 \rho_1}{\gamma (3H^2 - \rho_1)}\right]$ &  &  \\ \hline\hline
\end{tabular}%
\caption{The table presents different particle creation rates corresponding
to different effective fluids. }
\label{tab:characterization}
\end{table*}
\end{center}

\endgroup

\section{Singularities and analytic solutions}

\label{sec-analytic}

In this section we present the solutions for the master differential
equation (\ref{mas.01}) by applying the singularity analysis for a general
matter creation rate $\Gamma $. 
We begin our analysis by introducing the following generalized series for
the matter creation rate: 
\begin{equation}
\Gamma \left( H\right) =\Gamma _{0}+\Gamma _{1}H^{-1}+\Gamma
_{2}H^{-2}+\sum\limits_{i=3}^{n}\Gamma _{i}H^{-i},  \label{mas.02}
\end{equation}%
where the $\Gamma _{i}$'s ($i=0,1,...n$) are constants. From equation (\ref%
{mas.02}), we see that different choices of the constants, $\Gamma _{i}$,
give different matter creation rates. In particular, we can recover some
simplest matter creation rates, such as, $\Gamma =\Gamma _{0}$, $1/H$, $%
1/H^{2}$, as well as quadratic forms, $\Gamma =\Gamma _{0}+\Gamma _{1}/H$,
or $\Gamma _{0}+\Gamma _{2}/H^{2}$, and various others. In addition, we see
that in the early universe, where $H\rightarrow \infty $, it follows that $%
\Gamma \left( H\right) \simeq \Gamma _{0}$ from eqn. (\ref{mas.02}). \textbf{%
This is the focus of our investigation because we are looking for singular
solutions.}

The series (\ref{mas.02}) can also describe other forms of particle creation
function for specific values of the coefficients $\Gamma _{i}$; for
instance, when 
\begin{equation}
\Gamma _{j}=\frac{1}{j!}A^{j}\Gamma _{0}~,~j=1...n~\text{with }n\rightarrow
\infty
\end{equation}%
then, the series (\ref{mas.02}) is the Taylor expansion of the exponential
function%
\begin{equation}
\Gamma \left( H\right) =\Gamma _{0}e^{AH^{-1}}
\end{equation}%
around the point at which $H\rightarrow \infty $. Similarly, series (\ref%
{mas.02}) can describe other particle creation models for specific values of
the coefficients, and our analysis is valid for all the particle creation
models which can be described by the series (\ref{mas.02}), and hence, the
present work offers a general study of these particle creation models. The
solutions that we derive can also be seen as approximate or exact solutions
for all the particle creation models close to the singular solution which
corresponds to the dominant term in the series for $\Gamma (H)$.

\begin{figure}[ptb]
\includegraphics[height=13.5cm]{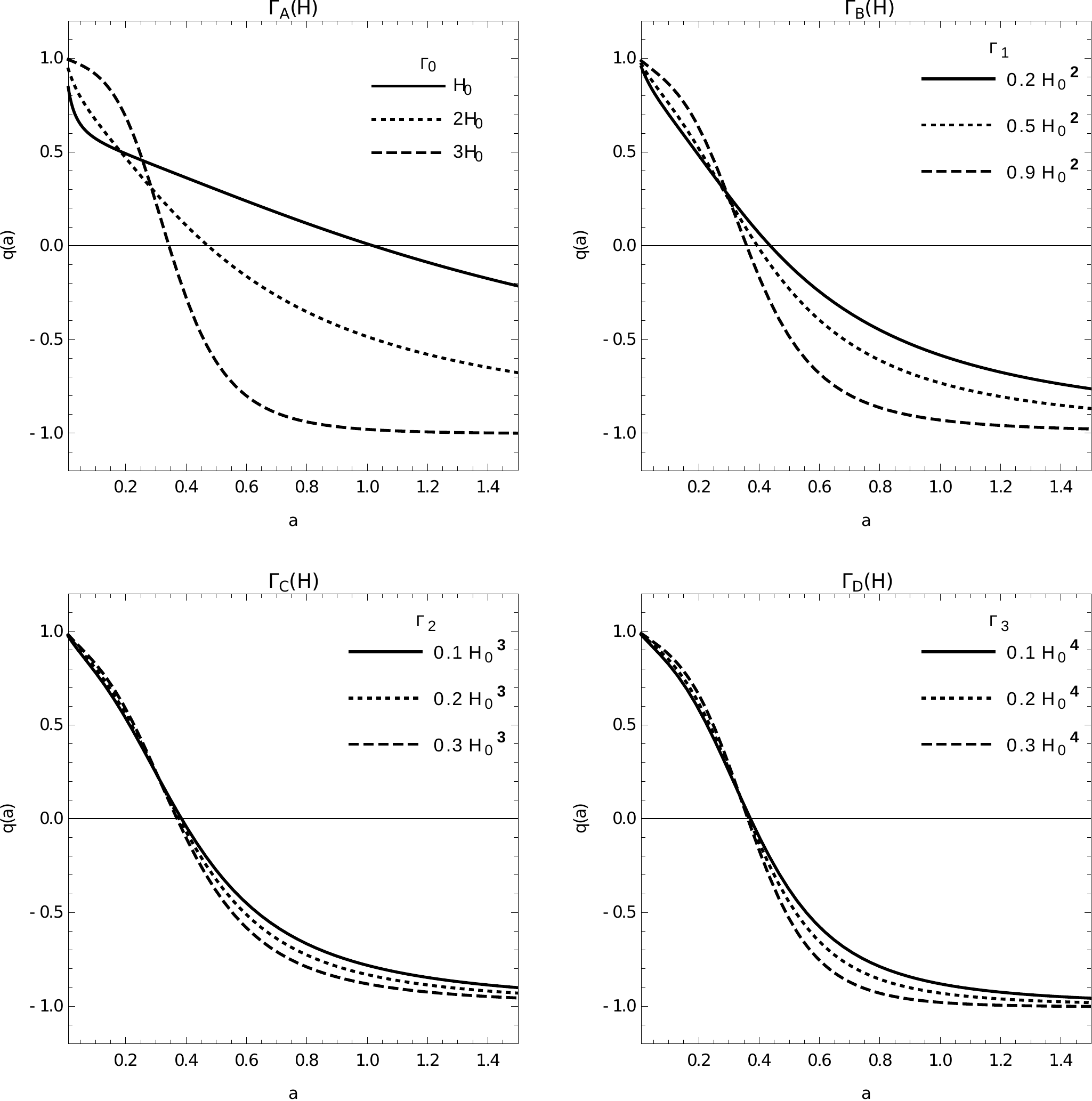}\centering
\caption{Qualitative evolution of the deceleration parameter $q\left(
a\right) $ given by the numerical solutions of equation (\protect\ref{mas.01}%
) for various particle creation models as given by the expression (\protect
\ref{mas.02}). The upper left panel stands for $\Gamma_A (H) =\Gamma_0$
where $\Gamma_0 = H_0$ (solid curve), $2 H_0$ (dashed curve) and $3 H_0$
(longdashed curve). The upper right panel for $\Gamma_B (H) = 2 H_0
+\Gamma_1 H^{-1} $ with $\Gamma_1 = 0.2 H_0^2$ (solid curve), $0.5 H_0^2$
(dashed curve), $0.9 H_0^2$ (longdashed curve). The lower left panel stands
for $\Gamma_C (H) = 2 H_0 + 0.5 H_0^2 H^{-1}+ \Gamma_2 H^{-2}$ with $%
\Gamma_2 = 0.1 H_0^3$ (solid curve), $0.2 H_0^3$ (dashed curve) and $0.3
H_0^3$ (longdashed curve). Finally, the lower right panel stands for $%
\Gamma_D (H) = 2 H_0 + 0.5 H_0^2 H^{-1}+ 0.2 H_0^3 H^{-2} + \Gamma_3 H^{-3}$
with $\Gamma_3 = 0.1 H_0^4$ (solid curve), $0.2 H_0^4$ (dashed curve) and $%
0.3 H_0^4$ (longdashed curve). The solutions are for $\protect\gamma=1$ and $%
\protect\gamma_{1}=\frac{4}{3}\,,$ and $\protect\rho_{1,0}=3%
\Omega_{r0}H_{0}^{2}$, with $\Omega_{r0}\simeq5\times10^{-3}$}
\label{fig1}
\end{figure}

\begin{figure}[ptb]
\includegraphics[height=13.5cm]{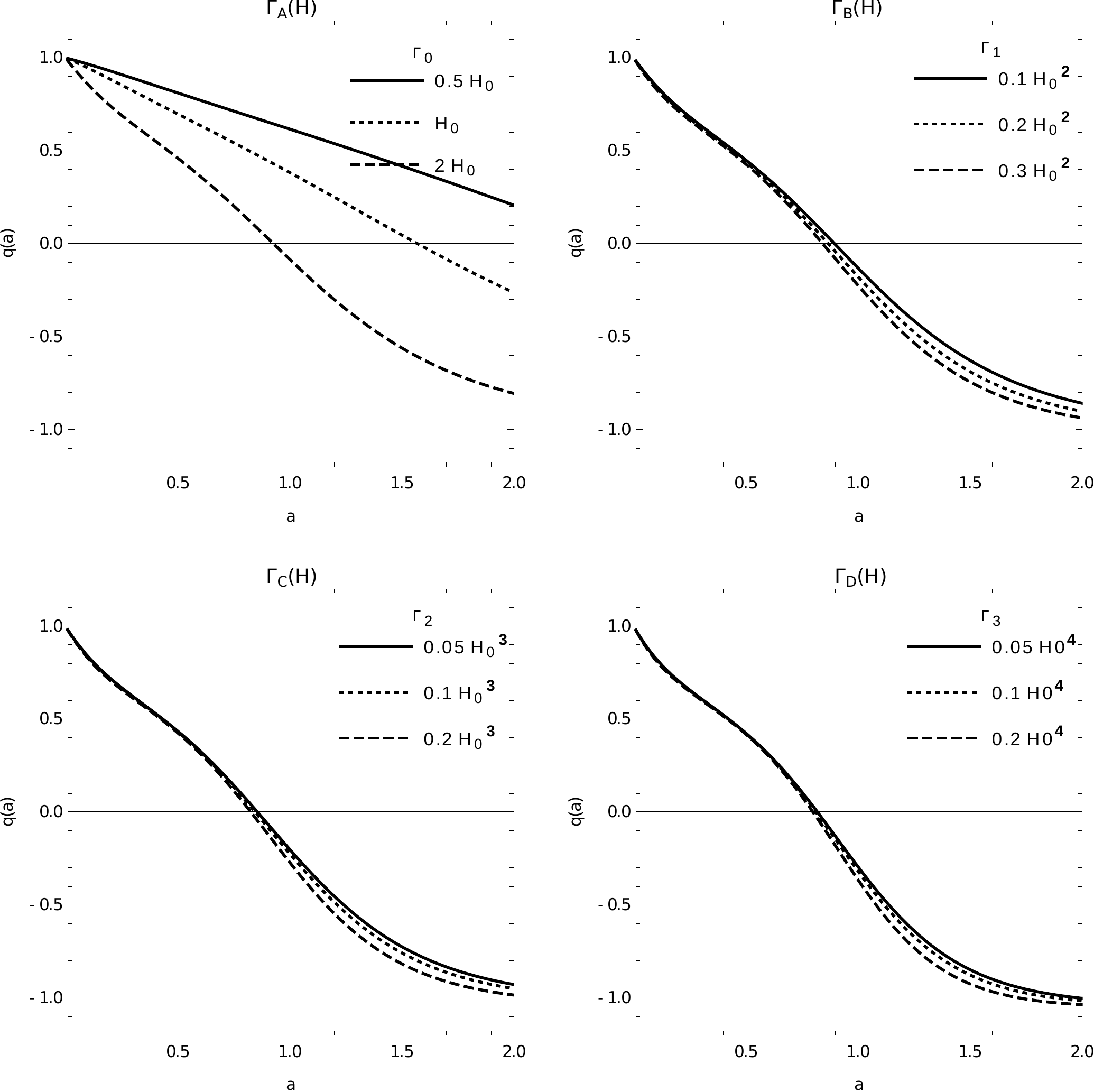}\centering
\caption{Qualitative evolution of the deceleration parameter $q\left(
a\right) $ given by numerical solutions of the master equation (\protect\ref%
{mas.01}) for various particle creation models as given by the expression (%
\protect\ref{mas.02}). The upper left panel stands for $\Gamma_A (H)
=\Gamma_0$ where $\Gamma_0 = 0.5 H_0$ (solid curve), $H_0$ (dashed curve)
and $2 H_0$ (longdashed curve). The upper right panel for $\Gamma_B (H) = 2
H_0 +\Gamma_1 H^{-1} $ with $\Gamma_1 = 0.1 H_0^2$ (solid curve), $0.2 H_0^2$
(dashed curve), $0.3 H_0^2$ (longdashed curve). The lower left panel stands
for $\Gamma_C (H) = 2 H_0 + 0.5 H_0^2 H^{-1}+ \Gamma_2 H^{-2}$ with $%
\Gamma_2 = 0.05 H_0^3$ (solid curve), $0.1 H_0^3$ (dashed curve) and $0.2
H_0^3$ (longdashed curve). Finally, the lower right panel stands for $%
\Gamma_D (H) = 2 H_0 + 0.5 H_0^2 H^{-1}+ 0.2 H_0^3 H^{-2} + \Gamma_3 H^{-3}$
with $\Gamma_3 = 0.05 H_0^4$ (solid curve), $0.1 H_0^4$ (dashed curve) and $%
0.2 H_0^4$ (longdashed curve). The solutions are for $\protect\gamma=\frac{4%
}{3}$ and $\protect\gamma_{1}=1$ with $\protect\rho_{1,0}=3%
\Omega_{m0}H_{0}^{2}$ and $\Omega_{m0}\simeq0.3$.}
\label{fig2}
\end{figure}

In order to understand the evolution of the cosmological model given by (\ref%
{mas.01}) for the prescribed particle creation rate (\ref{mas.02}) in
Figures \ref{fig1} and \ref{fig2}, we present the numerical evolution of the
deceleration parameter $q\left( a\right) $ for the initial conditions $%
a\left( t_{0}\right) =1$ and $\dot{a}\left( t_{0}\right) =70$, where the
present value of the Hubble constant is $H_{0}=\frac{\dot{a}\left(
t_{0}\right) }{a\left( t_{0}\right) }$.

The numerical solution in Fig. \ref{fig1} is for $\gamma =1$ and $\gamma
_{1}=\frac{4}{3}\,,$ and we select $\rho _{1,0}=3\Omega _{r0}H_{0}^{2}$,
with $\Omega _{r0}\simeq 5\times 10^{-3}$. \ On the other hand, Fig. \ref%
{fig2} is for $\gamma =\frac{4}{3}$ and $\gamma _{1}=1$ with $\rho
_{1,0}=3\Omega _{m0}H_{0}^{2}$ and $\Omega _{m0}\simeq 0.3$. In both figures
the $\Gamma _{i}$ coefficients of the particle creation function $\Gamma
\left( H\right) $ have been considered to be positive.

From Fig. \ref{fig1} and Fig. \ref{fig2}, we observe that in the early
universe, the universe is dominated by the matter source and there always
exists a de Sitter point as a future attractor. These de Sitter points can
be easily calculated in a similar way as in \cite{bar01}.

\subsection{ARS algorithm}

In order to derive the analytic solutions of the master equation (\ref%
{mas.01}) we work with the approach of Kowalevskaya \cite{Kowa}, and
determine if the second-order differential equation (\ref{mas.01}) possesses
the movable singularities. The existence of a movable singularity means that
the solution of equation (\ref{mas.01}) near the singularity is described by
the power-law function $a\left( \tau \right) \simeq \tau ^{p},~\tau
=t-t_{0}, $ where $p$ is a negative number and $t_{0}$ denotes the position
of the singularity. The position of the singularity changes according to the
initial condition of the problem which means that different initial
conditions provide us with different locations for the singular point.

Ablowitz, Ramani and Segur (ARS) \cite{Abl1,Abl2,Abl3}, proposed an
algorithm to test if a given differential equation is integrable when an
algebraic solution exists given by a Laurent expansion.

The ARS algorithm is briefly described by the following three main steps 
\cite{paperleach}:

(a) Determine the leading-order behaviour, at least in terms of the dominant
exponent. The coefficient of the leading-order term may or may not be
explicit.

(b) Determine the exponents at which the arbitrary constants of integration
enter.

(c) Substitute an expansion up to the maximum resonance into the full
equation to check for consistency.

For the singularity analysis 
the exponents of the leading-order term needs to be negative integers or a
non-integral rational number. However, this is not so restrictive since we
can always perform a change of coordinates in order to obtain a negative
exponent for the leading-order term.

In order that the differential equation passes the Painlev\'{e} test, the
resonances have to be rational numbers, so that the solution can be written
as a Painlev\'{e} series. Alternatively, if at least one of the resonances
is an irrational number, the differential equation passes the weak Painlev%
\'{e} test. Moreover, the value minus one $\left( -1\right) $ should always
appear as one of the resonances. The existence of that resonance is
important in order for the singularity to be movable. For a right Painlev%
\'{e} series (right Laurent expansion) the resonances must be nonnegative;
for a left Painlev\'{e} series (left Laurent expansion) the resonances must
be non-positive while for a full Laurent expansion the resonances have to be
mixed. Obviously,\ the possible Laurent expansions for second-order
differential equations are either left or right Painlev\'{e} series. For a
review and various applications we refer the reader to Ref. \cite{buntis},
while different applications of singularity analysis in cosmological studies
can be found in \cite{Paliathanasis:2016vsw,sin1,sin2,sin3,sin3b,sin4,sin5}
and references therein.

\subsection{Leading-order behaviour}

We continue by applying the first step of the ARS algorithm to determine the
leading-order behaviour. We substitute, $a\left( t\right) =a_{0}\left(
t-t_{0}\right) ^{p}$, in the master equation (\ref{mas.01}), where we find
that for $\gamma \neq \gamma _{1}$, there are two possible leading-order
behaviours are described by the power-law functions\footnote{%
The leading-behaviour $a_{B}\left( t\right) $ exists only when $\gamma \geq
\gamma _{1}.$}%
\begin{equation}
a_{A}\left( \tau \right) =a_{A0}\tau ^{\frac{2}{3\gamma _{1}}}\text{, and }%
a_{B}\left( \tau \right) =a_{B0}\tau ^{\frac{2}{3\gamma }}.
\end{equation}

The scale factors $a_{B}\left( \tau \right) $ describe solutions with
perfect fluids. The function $a_{A}\left( \tau \right) $ corresponds to the
matter solutions in which the perfect fluid with $\gamma _{1}$ dominates,
while function $a_{B}\left( \tau \right) \ $describes a solution in which
the fluid term with $\gamma $ dominates.

Furthermore, the coefficient parameter $a_{B0}$ is found to be arbitrary,
while $a_{A0}$ is related to the energy density $\rho _{1,0}$ by%
\begin{equation}
\rho _{1,0}=\frac{4a_{A0}^{3\gamma _{1}}}{3\left( \gamma _{1}\right) ^{2}}%
\text{.}
\end{equation}%
The coefficient $a_{B0}$ is the second-integration constant which controls
the generic solution of the master equation (\ref{mas.01}). The position in
the Laurent expansion of the second integration constant in the
leading-order behaviour, $a_{A0}$, is determined by the values of the
resonances. The later will be found below. Finally, note that in the case
where $\gamma =\gamma _{1}$, there exists only one leading-order term, the $%
a_{B}\left( t\right) ,$ with an arbitrary value for the coefficient $a_{B0}$.

\subsection{Resonances}

The second step in the\ ARS algorithm is the determination of the
resonances. We substitute the following expression,%
\begin{equation}
a\left( \tau \right) =a_{\left( A,B\right) }\left( \tau \right) \left(
1+\varepsilon \tau ^{s}\right) ,
\end{equation}%
in the master equation (\ref{mas.01}) then we linearize around $\varepsilon
=0.$ The coefficients of the leading-order behaviour provide a polynomial
equation of order two, whose solutions are the resonances, $s$, of the
leading-order behaviours $a_{A}\left( \tau \right) $ and $a_{B}\left( \tau
\right) $.

For the dominant term, $a_{A}\left( t\right) ,$ the resonances are
calculated to be%
\begin{equation}
s_{A1}=-1~,~s_{A2}=-\frac{2}{\gamma _{1}}\left( \gamma -\gamma _{1}\right) .
\end{equation}%
This means that, for $\gamma >\gamma _{1}$,$~$the solution will be given by
a left Painlev\'{e} series, while when $\gamma <\gamma _{1}$~the solution is
given by a right Painlev\'{e} series.

For the dominant term, $a_{B}\left( t\right) $, the resonances are
calculated to be%
\begin{equation}
s_{B1}=-1~,~s_{B2}=0.
\end{equation}%
Resonance $s_{B2}$ indicates that the second integration constant is $a_{B0}$
-- which we calculated above.

In the case where $\gamma =\gamma _{1}$, the two resonances are found to be%
\begin{equation}
s_{C1}=-1~,~s_{C2}=0.
\end{equation}

From the values of these resonances we can extract more information than the
position of the second integration constant. In particular, from the
discussion in \cite{anleach} we find that when the solution is given by a
left Painlev\'{e} series, the leading-order term describes an attractor
solution, while when the solution is given by a right Painlev\'{e} series
the leading-order behaviour is described by a source point and so is an
unstable solution. Furthermore, when the solution is given by a full Laurent
expansion the leading-order behaviour is described by a saddle point.

\subsection{Consistency tests and analytic solutions}

We continue with the consistency test and we write the analytic solution in
terms of Laurent expansions. However, in order to continue it is necessary
to select values of the equation of state parameters of the two perfect
fluids.

The consistency tests that we present are for combinations of dust and
radiation:

(A) $\left( \gamma ,\gamma _{1}\right) =\left( 1,\frac{4}{3}\right) $,~(B)~$%
\left( \gamma ,\gamma _{1}\right) =\left( \frac{4}{3},1\right) $,~(C)$%
~\gamma =\gamma _{1}$ with $\gamma =\frac{4}{3},~$and~(D) $\gamma =\gamma
_{1}$ with $\gamma =1$.

The coefficient terms that we calculate are the \textit{first} terms until
the parameter $\Gamma _{2}$ of (\ref{mas.02}) appears.

\subsubsection{Dust with particle creation and radiation}

When $\left( \gamma ,\gamma _{1}\right) =\left( 1,\frac{4}{3}\right) $, the
resonances which correspond to the leading-order term $a_{A}\left( \tau
\right) =a_{A0}\tau ^{\frac{1}{2}},$ take the values%
\begin{equation}
s_{A1}=-1~,~s_{A2}=\frac{1}{2}.
\end{equation}%
This means that the solution starts from the radiation era and is given by a
right Painlev\'{e} series. Moreover, we can infer that the solution in the
radiation era is a source, that is, it is an unstable solution.

The analytic solution is given by the right Painlev\'{e} series with step $%
\frac{1}{2}$, 
\begin{equation*}
\frac{a_{A}\left( \tau \right) }{a_{A0}}=\tau ^{\frac{1}{2}}+a_{A1}\tau
+a_{A2}\tau ^{\frac{3}{2}}+\sum\limits_{i=3}^{\infty }a_{Ai}\tau ^{\frac{1+i%
}{2}}
\end{equation*}%
in which $\rho _{1,0}=\frac{4}{3}\left( a_{A0}\right) ^{3}$.

The second integration constant of the solution is the coefficient term $%
a_{A1}$, while the rest of the coefficients, $a_{Ai}$, are functions of $%
a_{A1}$ and the parameters $\Gamma _{i}$, as follows:%
\begin{equation*}
a_{A2}=-\frac{7}{8}\left( a_{A1}\right) ^{2}\,,~a_{A3}=\frac{5}{4}\left(
a_{A1}\right) ^{3}+\frac{3}{5}a_{A1}\Gamma _{0},
\end{equation*}%
\begin{equation*}
a_{A4}=-\frac{273}{128}\left( a_{A1}\right) ^{4}-\frac{13}{10}\left(
a_{A1}\right) ^{2}\Gamma _{0},
\end{equation*}%
\begin{equation*}
a_{A5}=4\left( a_{A1}\right) ^{5}+\frac{467}{140}\left( a_{A1}\right)
^{3}\Gamma _{0}+\frac{3}{14}a_{A1}\left( \left( \Gamma _{0}\right)
^{2}+2\Gamma _{1}\right) ,
\end{equation*}%
\begin{equation*}
a_{A6}=-\frac{8151}{1024}\left( a_{A1}\right) ^{6}-\frac{2007}{224}\left(
a_{A1}\right) ^{4}\Gamma _{0}-\frac{3}{1400}\left( a_{A1}\right) ^{2}\left(
479\left( \Gamma _{0}\right) ^{2}+615\Gamma _{1}\right) ,
\end{equation*}%
and%
\begin{equation*}
a_{A7}=\frac{33}{2}\left( a_{A1}\right) ^{7}+\frac{a_{A1}\Gamma _{0}}{720}%
\left( 17775\left( a_{A1}\right) ^{4}+3174\left( a_{A1}\right) ^{2}\Gamma
_{0}+40\left( \Gamma _{0}\right) ^{2}\right) +\frac{22}{5}\left(
a_{A1}\right) ^{3}\Gamma _{1}+\frac{a_{A1}}{9}\left( 3\Gamma _{0}\Gamma
_{1}+4\Gamma _{2}\right) .
\end{equation*}

From these expressions it is clear that the higher polynomial terms of the
particle creation function (\ref{mas.02}) play a crucial role in the
analytic solution as we evolve far from the radiation era.

\subsubsection{Radiation with Particle creation and pressureless fluid}

In order to test the consistency of the second solution with leading-order
term $a_{B}\left( \tau\right) $, we select $\gamma=\frac{4}{3}$ and $%
\gamma_{1}=1$.

Hence, the analytic solution is given by the right Painlev\'{e} series with
step $\frac{1}{2}$, 
\begin{equation*}
\frac{a_{B}\left( \tau \right) }{a_{B0}}=\tau ^{\frac{1}{2}}+a_{B1}\tau
+a_{B2}\tau ^{\frac{3}{2}}+\sum\limits_{i=3}^{\infty }a_{Bi}\tau ^{\frac{1+i%
}{2}},
\end{equation*}%
where now $a_{B0}$ is the second integration constant. The first six
coefficients are given by the following expressions%
\begin{equation*}
a_{B1}=\frac{2\rho _{1,0}}{\left( a_{B0}\right) ^{3}}~,~a_{B2}=-\frac{7}{8}%
\left( a_{B1}\right) ^{2}+\frac{\Gamma _{0}}{6}~~,~a_{B3}=\frac{5}{4}\left(
a_{B1}\right) ^{3}-\frac{7}{15}a_{B1}\Gamma _{0},
\end{equation*}%
\begin{equation*}
a_{B4}=-\frac{273}{128}\left( a_{B1}\right) ^{4}+\frac{31}{240}\left(
a_{B1}\right) ^{2}\Gamma _{0}+\frac{5}{216}\left( \Gamma _{0}\right) ^{2}+%
\frac{1}{9}\Gamma _{1},
\end{equation*}%
\begin{equation*}
a_{B5}=5\left( a_{B1}\right) ^{5}-\frac{253}{70}\left( a_{B1}\right)
^{3}\Gamma _{0}+\frac{106}{945}a_{B1}\left( \Gamma _{0}\right) ^{2}-\frac{134%
}{315}a_{B1}\Gamma _{1},
\end{equation*}%
and%
\begin{equation*}
a_{B6}=-\frac{8151}{1024}\left( a_{B1}\right) ^{6}+\frac{18223}{1792}\left(
a_{B1}\right) ^{4}\Gamma _{0}+\left( a_{B1}\right) ^{2}\left( \frac{213}{140}%
\Gamma _{1}-\frac{97411}{100800}\left( \Gamma _{0}\right) ^{2}\right) +\frac{%
\left( \Gamma _{0}\right) ^{3}+8\Gamma _{0}\Gamma _{1}+48\Gamma _{2}}{432}.
\end{equation*}%
Again, we observe that as we go far from the leading-order term, i.e. from
the radiation era, the coefficients of the higher polynomial terms are
involved in the solution.

\subsubsection{Radiation-like fluid with particle creation and radiation}

Now, in the case in which $\gamma =\gamma _{1}$ and $\gamma =\frac{4}{3}$,
the generic analytic solution is given by the Laurent expansion%
\begin{equation*}
\frac{a_{C}\left( \tau \right) }{a_{C0}}=\tau ^{\frac{1}{2}}+a_{C1}\tau
+a_{C2}\tau ^{\frac{3}{2}}+\sum\limits_{i=3}^{\infty }a_{Ci}\tau ^{\frac{1+i%
}{2}},
\end{equation*}%
where $a_{C0}$ is arbitrary and the first non-zero coefficients are 
\begin{equation*}
a_{C2}=\frac{9}{32}+\frac{\Gamma _{0}}{6}~,~a_{C4}=\frac{\left( 27+16\Gamma
_{0}\right) \left( \Gamma _{0}\left( 80\Gamma _{0}-297\right) +384\Gamma
_{1}\right) }{55296\Gamma _{0}},
\end{equation*}%
\begin{equation*}
a_{C6}=\frac{27+16\Gamma _{0}}{1769472\Gamma _{0}}\left( \Gamma _{0}\left(
8505+32\Gamma _{0}\left( 27+8\Gamma _{0}\right) +2048\Gamma _{1}\right)
-384\left( 45\Gamma _{1}-32\Gamma _{2}\right) \right) .
\end{equation*}%
We note that the non-zero coefficients are the $a_{Ci}=a_{C\left( 2k\right)
} $, $~k\in 
\mathbb{N}
.$

\subsubsection{Dust with particle creation and pressureless fluid}

In a similar way, when $\gamma =\gamma _{1}$ and $\gamma =1$, the general
analytic solution is expressed by the Laurent expansion%
\begin{equation*}
\frac{a_{D}\left( \tau \right) }{a_{D_{0}}}=\tau ^{\frac{2}{3}}+a_{D1}\tau
+a_{D2}\tau ^{\frac{4}{3}}+\sum\limits_{i=3}^{\infty }a_{Di}\tau ^{\frac{2+i%
}{3}},
\end{equation*}%
which starts from the matter-dominated era.

The coefficient $a_{D0}$ is the second integration constant of the solution,
while the first non-zero coefficients are calculated to be,%
\begin{equation*}
a_{D3}=\frac{3}{8}+\frac{\Gamma _{0}}{6}~,~a_{D6}=\frac{\left( 9+4\Gamma
_{0}\right) }{769\Gamma _{0}}\left( \Gamma _{0}\left( 4\Gamma _{0}-15\right)
+16\Gamma _{1}\right) ,
\end{equation*}%
\begin{equation*}
a_{D9}=\frac{9+4\Gamma _{0}}{165888\Gamma _{0}}\left( \Gamma _{0}\left(
16\Gamma _{0}\left( 9+5\Gamma _{0}\right) +9\left( 315+64\Gamma _{1}\right)
\right) -648\left( 7\Gamma _{1}-4\Gamma _{2}\right) \right) .
\end{equation*}%
We note that the non-zero coefficients are the $a_{Di}=a_{D\left( 3k\right)
} $, $~k\in 
\mathbb{N}
.$

\section{Conclusions}

\label{discu}

\emph{\ }An explicit form of dark energy or the presence of modifications to
general relativity gravity are two independent roads towards an explanation
for the observed acceleration of the universe. Here we consider a third
alternative through which we can describe the observational results in a
convenient way. A theory of gravitational particle production has shown that
different phases of the universe can be explained \cite{Prigogine-inf,
Abramo:1996ip, Gunzig:1997tk,Zimdahl:1999tn, SSL09, LJO10, LB10, JOBL11,
LBC12, LGPB14, CPS14, FPP14, NP15, lima16, NP16, HP16, RCN16, PHPS16} and so
it was argued that the particle creation formalism might be considered as a
viable alternative to the dark energy or modified gravitational theories.
The key role in gravitational particle production is played by the rate of
particle creation $\Gamma (t)$ which has an equivalent character to an
effective dynamical equation of state, and so we can consider various
phenomenological models for $\Gamma (t)$. But, except for some simple
choices of the creation rate, the dynamics of the universe cannot be
obtained in an analytic way. This motivated the present paper, where for the
first time we present the exact solutions of the gravitational field
equations for a system of two fluids model including particle creation.

In particular, assuming a spatially flat Friedmann-Lema\^{\i}%
tre-Robertson-Walker universe described by the Einstein gravity, we consider
two perfect fluids with barotropic equations of state where one of them is
endowed with particle creation, as if it possessed a bulk viscosity, while
the other fluid obeys the standard conservation law. Interestingly, we found
that the dynamics of such a two-fluid particle creation system can be
concisely described by a single nonlinear differential equation (\ref{mas.01}%
) that involves the particle creation rate $\Gamma (t)$. We recall that the
above two-fluid particle creation system might be considered to be an
equivalent cosmological scenario to a single fluid associated with particle
creation in the presence of curvature \cite{bar01}. However, since the
particle creation rate $\Gamma (t)$ is any unknown function and its detailed
functional form is unknown (it must be derived from QFT in future), we widen
our investigations by allowing a general particle creation rate (\ref{mas.02}%
) that recovers some specific particle creation models as special cases,
namely, $\Gamma =\Gamma _{0}=$ constant, $\Gamma \propto 1/H^{n}$ ($n\in 
\mathbb{N}$), $\Gamma =span\{\Gamma _{0},H^{-1},H^{-2},...,H^{-n}\}$ ($n\in 
\mathbb{N}$), as well as some other exceptional but interesting choices like 
$\Gamma \propto \exp (1/H)$. Following a singularity analysis applied to the
nonlinear differential equation defining the theory, we see that the master
equation in this work, i.e., equation (\ref{mas.01}) can pass the
singularity test and hence, the solution for the gravitational equations can
be written in terms of the Laurent series around the movable singularity. We
note that we do not consider choices like $\Gamma \propto H^{2}$ in the
general model of $\Gamma $ in (\ref{mas.02}), since for such models, the
governing differential equation does not pass the singularity test \cite%
{bar01}.

Therefore we see that for the present two-fluid cosmological model with
gravitationally induced \textquotedblleft adiabatic\textquotedblright\
particle creation, we can obtain the singular algebraic solutions to the
gravitational field equations for a class of particle creation models given
in (\ref{mas.01}). Finally, we mention that the present work can be extended
to more than two fluids model in presence of the particle creation, although
the dynamics could be more complicated .

\begin{acknowledgments}
The authors thank the referee for some specific comments on the article.  SP
acknowledges the research grant under Faculty Research and Professional
Development Fund (FRPDF) Scheme of Presidency University, Kolkata, India.
JDB was supported by the Science and Technology Facilities Council (STFC) of
the United Kingdom. AP acknowledges the financial support \ of FONDECYT
grant no. 3160121.
\end{acknowledgments}


\end{document}